\documentclass[a4paper,twocolumn]{revtex4}
\usepackage{color}
\usepackage{graphics}
\usepackage{epsfig}
\usepackage{natbib}

\begin{document}

\title{Emergence of weak pyrochlore phase and signature of field induced spin ice ground state in Dy$_{2-x}$La$_{x}$Zr$_{2}$O$_{7}$; x = 0, 0.15, 0.3}

\author{Sheetal$^{1}$, Anzar Ali$^{2}$, Sarita Rajput$^{3}$, Yogesh Singh$^{2}$, T.Maitra$^{3}$, and C.S. Yadav$^{1*}$}
\affiliation{$^{1}$School of Basic Sciences, Indian Institute of Technology Mandi, Mandi-175005, (H.P.), India}
\affiliation{$^{2}$Department of Physics, Indian Institute of Science Education and Research Mohali, Mohali-140306 , (Punjab), India}
\affiliation{$^{3}$Department of Physics, Indian Institute of Technology Roorkee, Roorkee-247667, (Uttrakhand), India}

\begin{abstract}
The pyrochlore oxides Dy$_{2}$Ti$_{2}$O$_{7}$ and Ho$_{2}$Ti$_{2}$O$_{7}$ are well studied spin ice systems and have shown the evidences of magnetic monopole excitations. Unlike these, Dy$_{2}$Zr$_{2}$O$_{7}$ is reported to crystallize in a distorted fluorite structure. We present here the magnetic and heat capacity studies of La substituted Dy$_{2}$Zr$_{2}$O$_{7}$. Our findings suggest the absence of spin ice state in Dy$_{2}$Zr$_{2}$O$_{7}$ but the emergence of the magnetic field induced spin freezing near T $\approx$ 10 K in ac susceptibility measurements which is similar to Dy$_{2}$Ti$_{2}$O$_{7}$. The magnetic heat capacity of Dy$_{2}$Zr$_{2}$O$_{7}$ shows a shift in the peak position from 1.2 K in zero field to higher temperatures in the magnetic field, with the corresponding decrease in the magnetic entropy. The low temperature magnetic entropy at 5 kOe field is Rln2 - (1/2)Rln(3/2) which is same as for the spin ice state. Substitution of non-magnetic, isovalent La$^{3+}$ for Dy$^{3+}$ gradually induces the structural change from highly disordered fluorite to weakly ordered pyrochlore phase. The La$^{3+}$ substituted compounds with less distorted pyrochlore phase show the spin freezing at lower field which strengthens further on the application of magnetic field. Our results suggest that the spin ice state can be stabilized in Dy$_{2}$Zr$_{2}$O$_{7}$ either by slowing down of the spin dynamics or by strengthening the pyrochlore phase by suitable substitution in the system.
\end{abstract}

\maketitle

\section{Introduction}

Geometrically frustrated magnetic systems have been a subject of continuing research interest because of the realization of interesting quantum phases, non-trivial magnetic ordering and excitations originating from the competing ferromagnetic (FM) and antiferromagnetic (AFM) interactions \cite{misguich2005frustrated, castelnovo2008magnetic, gardner2010magnetic, bovo2014restoration, balents2010spin,snyder2002dirty}. Of particular interest are the pyrochlore oxides: Dy$_{2}$Ti$_{2}$O$_{7}$, Nd$_{2}$Zr$_{2}$O$_{7}$ and Ho$_{2}$Ti$_{2}$O$_{7}$ where the exotic magnetic and thermodynamic properties have been observed \cite{matsuhira2001novel,bramwell2001spin,anand2015investigations, anand2015observation} at low temperature. Pyrochlore oxide A$_{2}$B$_{2}$O$_{7}$, (A is the trivalent rare-earth ion and B tetravalent transition metal ion) contains a network of corner-sharing tetrahedra where the magnetic rare-earth ions reside at the corners of tetrahedra and play a major role in deciding the magnetic behavior of these systems \cite{siddharthan1999ising,giauque1936entropy}. In pyrochlores, the exotic magnetic ground state is achieved by three competing interactions: AFM exchange interaction, FM dipolar interactions and crystal electric field (CEF), where the dominant effect of CEF forces the spin to point either directly towards or away from the center of the tetrahedra and induces spin anisotropy in the system. For example, in Ho$_{2}$Ti$_{2}$O$_{7}$ and Dy$_{2}$Ti$_{2}$O$_{7}$ spin ice ground state is achieved by spins and their excitations are magnetic monopoles, Er$_{2}$Ti$_{2}$O$_{7}$ has the AFM ordering which is achieved through order-by-disorder mechanism and spin liquid ground state is observed in Tb$_{2}$Ti$_{2}$O$_{7}$ \cite{harris1997geometrical, sen2015topological, ross2014order, takatsu2011quantum}. The computational studies on these systems have suggested the presence of spin ice state (two in-two out ordering) due to the dominance of FM dipolar interactions over the AFM exchange interactions which favors all-in-all-out ordering. \cite{takatsu2013ac,melko2004monte}

More recently Ramon \textit{et al.} have shown Dy$_{2}$Zr$_{2}$O$_{7}$ to crystallize in defect-fluorite structure with lattice constant a = 5.238(2) $\AA$ \cite{ramon2019absence}. Unlike Dy$_{2}$Ti$_{2}$O$_{7}$, Dy$_{2}$Zr$_{2}$O$_{7}$ shows the absence of spin ice state \cite{ramirez1999zero}. Although the specific heat and neutron scattering studies suggest very dynamic, short range, AFM spin-spin correlation below 10 K and the system remains disordered down to 40 mK with a significant value of magnetic susceptibility \cite{ramon2019absence}. In Dy$_{2}$Ti$_{2}$O$_{7}$, the spin state satisfy the two-in two-out ice rule and possess the Pauling's residual entropy of (R/2)ln(3/2), which is absent in case of Dy$_{2}$Zr$_{2}$O$_{7}$ \cite{ramirez1999zero, ramon2019absence}. However, similar to Dy$_{2}$Ti$_{2}$O$_{7}$ a correlation peak is present in Dy$_{2}$Zr$_{2}$O$_{7}$ around 2 K in the heat capacity measurement \cite{ramon2019absence}. These results indicate that the combined effect of Zr$^{4+}$ disorder and spin frustration at low temperature leads to the dynamic ground state of Dy$_{2}$Zr$_{2}$O$_{7}$ \cite{ramon2019absence}. 

Here, we present the structural, magnetic and thermodynamic properties of Dy$_{2-x}$La$_{x}$Zr$_{2}$O$_{7}$; x = 0, 0.15, 0.3 based on dc magnetic susceptibility, ac magnetic susceptibility, isothermal magnetization and heat capacity measurements. It is observed that Dy$_{2}$Zr$_{2}$O$_{7}$ is a frustrated magnetic system which undergoes spin freezing transition at T$_{f}$ $\sim$ 10 K in ac susceptibility measured in the presence of dc magnetic field of 5 kOe. This spin freezing transition is similar to the transition in Dy$_{2}$Ti$_{2}$O$_{7}$ at T = 16 K in zero dc field, which is lost due to large structural disorder introduced by Zr atoms, and recovers on application of magnetic field. The partial substitution of non-magnetic La$^{3+}$ in place of magnetic Dy$^{3+}$ stabilizes the pyrochlore phase and spin-ice behavior is seen at lower fields. These results suggests that magnetic field can induce the spin-ice state in Dy$_{2}$Zr$_{2}$O$_{7}$ and La substitution can favor spin-ice formation at even lower fields.

\section{Experimental Details}

The polycrystalline Dy$_{2-x}$La$_{x}$Zr$_{2}$O$_{7}$ (x = 0, 0.15, 0.3) compounds were prepared by the reaction of stoichiometric mixture of constituent oxides: Dy$_{2}$O$_{3}$ (Sigma Aldrich, $\geq$99.99$\%$ purity), La$_{2}$O$_{3}$ (Sigma Aldrich, $\geq$99.999$\%$ purity) and ZrO$_{2}$ (Sigma Aldrich, 99$\%$ purity) in the alumina crucible at 1350$^{o}$C in air for 50 hours \cite{snyder2001spin}. The reaction at this temperature was done thrice with the intermediate grindings using an agate mortar and pestle. The Dy$_{2}$O$_{3}$ and La$_{2}$O$_{3}$ were pre-heated at 500$^{o}$C to get rid of any possible moisture because of the hygroscopic nature of rare earth oxides. The obtained compounds were further pelletized and sintered at 1350$^{o}$C for 50 hours. 

The crystal structure and phase quality of the sample was confirmed by the Rietveld refinement of the x-ray diffraction (XRD) pattern using Fullprof Suit software. Raman spectra of the compound was obtained at 300 K in back scattering geometry by using Horiba HR-Evolution spectrometer with 532 nm excitation laser. The x-ray photoemission spectroscopy (XPS) showed the Dy and Zr atoms to exhibit +3 and +4 oxidation state respectively (shown in figure 3), similar to Dy and Ti in Dy$_{2}$Ti$_{2}$O$_{7}$ and XPS spectrum does not show any peak corresponding to the oxygen vacancy \cite{snyder2002dirty}. The magnetic measurements were performed using Quantum Design built Magnetic Property Measurement System (MPMS) and Heat capacity was measured using Quantum Design built Physical Property Measurement System (PPMS).

We performed DFT calculations using the projector-augmented wave (PAW) psuedopotential and a plane wave based method as implemented in the Vienna Ab-initio Simulation Program \cite{kresse1996efficient}. The exchange-correlation functional used in our calculations is Perdew-Burke-Ernzerhof generalized gradient approximation (PBE-GGA) \cite{perdew1996generalized}. Coulomb interaction (U) and spin-orbit interaction (SO) were considered within GGA+U+SO approximation \cite{anisimov1993density}. An energy cut-off of 500 eV was used for the plane waves in the basis set while a 5$\times$5$\times$5 Monkhorst-Pack {\bf k}-mesh centered at ${\Gamma}$ was used for performing the Brillouin zone integrations.

\section{Results and Discussions}

\subsection{Crystal Structure}

\begin{figure}[htbp]
	\begin{center}
		\vspace{-10pt}
		\includegraphics[scale = 0.5]{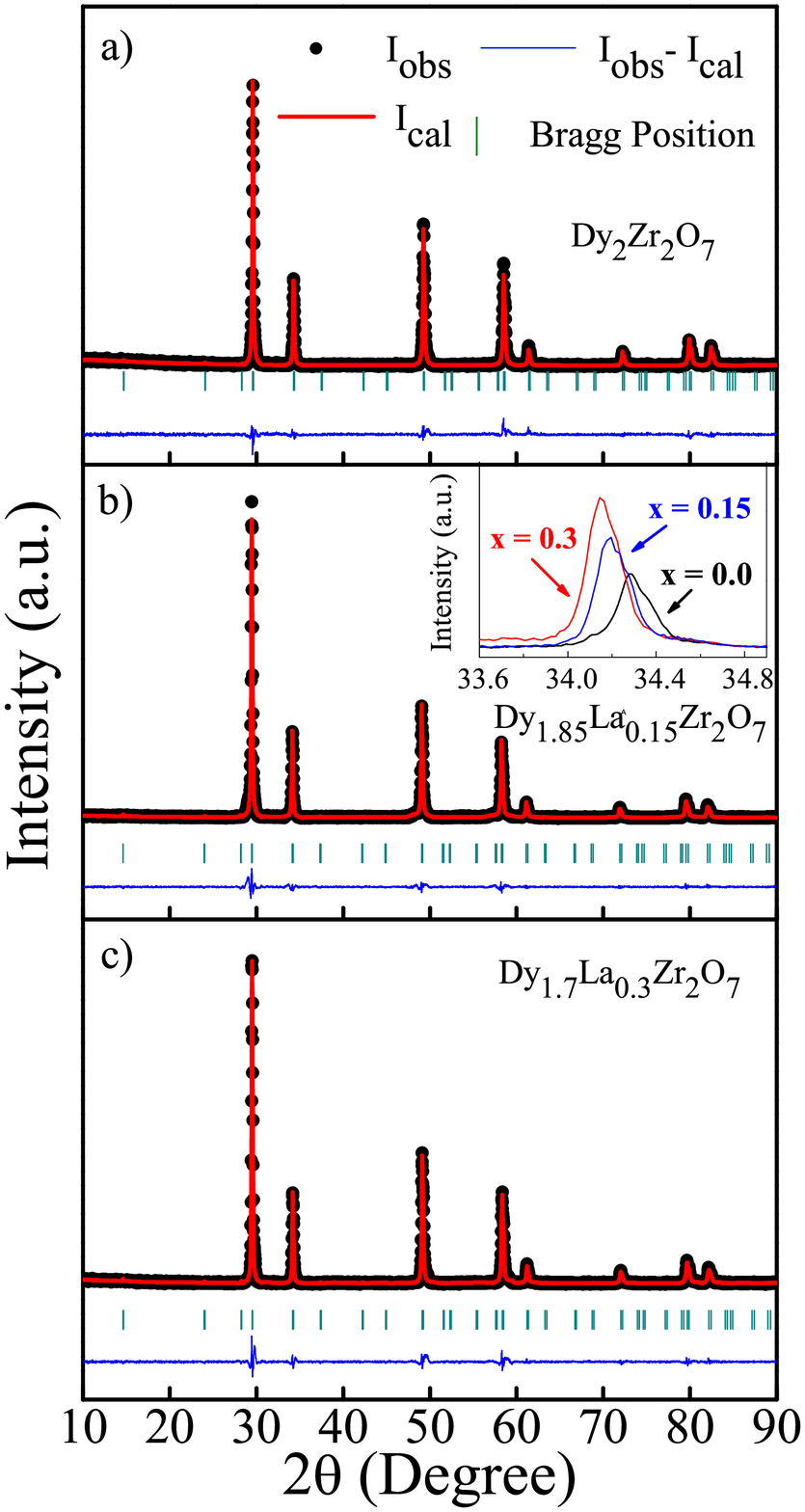}
		\vspace{-40pt}
		\caption{ X-ray diffraction pattern of Dy$_{2-x}$La$_{x}$Zr$_{2}$O$_{7}$ for x = 0.0, 0.15 and 0.3. collected at room temperature along with Rietveld refinement using Fd$\bar{3}$m space group. Insets of 1b and 1c represents the shift in peak position, lattice constant and x(O) parameter for the compounds respectively.}
	\end{center}
	\vspace{-20pt}
\end{figure}

Figure 1a shows the Rietveld refined XRD data of Dy$_{2}$Zr$_{2}$O$_{7}$ at T = 300 K fitted with space group Fd$\bar{3}$m. The XRD pattern consists of main peaks belonging to the pyrochlore lattice but some of the super-structural peaks corresponding to 2$\theta$ = 14$^{o}$ (111), 27$^{o}$ (311), 36$^{o}$ (331), 42$^{o}$ (422) are missing \cite{mandal2006preparation}. The XRD of Dy$_{2-x}$La$_{x}$Zr$_{2}$O$_{7}$; x = 0.15, 0.3 (figure 1b and 1c) also shows similar patterns with a slight shift in peak position towards lower angle (shown in the inset of figure 1b). The variation in lattice constant and position of oxygen ion x(O) with La substitution is plotted in the inset of figure 1c. Recently Ramon \textit{et al.} has reported a good fit for Dy$_{2}$Zr$_{2}$O$_{7}$ with Fm$\bar{3}$m disordered fluorite space group \cite{ramon2019absence}. However we have found a better fit of XRD data with Fd$\bar{3}$m space group ($\chi^{2}$ = 1.46) compared to Fm$\bar{3}$m ($\chi^{2}$ = 1.86).

In pyrochlore structure, 1/8 of the anion sites are vacant and these vacancies occupy a distinct site. Whereas in fluorite structure the O atoms locally form a perfect cube around both A and B atoms which is different from the configuration of oxygen atom in ideal pyrochlore structure. This symmetric rearrangement of O about the magnetic ion modifies the crystal field states expected for the pyrochlore structure \cite{mandal2006preparation, clancy2016x}. The transition from fluorite to pyrochlore phase arises due to ordered oxygen vacancy which results in the decrease in x-coordinate of 48f oxygen site. The variation in x(O) 48f site strongly affects the structural symmetry and determine the shape of polyhedra around A and B site \cite{hozoi2014longer,kumar2019spin}.

\begin{table}[htbp]
	\caption{Crystallographic data and magnetic parameters of Dy$_{2-X}$La$_{x}$Zr$_{2}$O$_{7}$; x = 0, 0.15, 0.3}
	\begin{tabular}{ccccccccccccc}
		\hline
		&&&& x= 0 &&&& x = 0.15 &&&& x = 0.3 \\
		\hline
		a ($\AA$)&&&&  10.4511(3) &&&& 10.4832(4) &&&& 10.4972(3)\\
		x(O) 	&&&&  0.3662 &&&& 0.3674 &&&&  0.3692 \\
		r$_{A}$/r$_{B}$ 	&&&&  1.43 &&&& 1.44 &&&& 1.45\\
		$\chi^{2}$ &&&& 1.46 &	&	& & 2.26 &&&& 2.22 \\
		$\mu_{eff} (\mu_{B})$ 	&&&& 8.20 &&&& 7.99 &&&& 7.06 \\
		$\theta_{cw} (K) $ &&&& -9.6 (2) &&&& -7.91 (2) &&&& -7.14 (1) \\
		J$_{nn}$/D$_{nn}$ &&&& -0.86 &&&& -0.863 &&&& -0.87 \\
		\hline
	\end{tabular}
\end{table}

For x$_{ideal}$ = 0.3125, B sites form an ideal octahedron and position of x = 0.375 indicates distorted octahedron for B atom and perfect cubic symmetry for A atom. The value of x(O) for Dy$_{2}$Ti$_{2}$O$_{7}$ is 0.4243 and replacement of Ti with Zr reduces it to 0.3662 for Dy$_{2}$Zr$_{2}$O$_{7}$ \cite{anand2015investigations}. This large variation in x(O) with Zr substitution expectantly changes the coordination of B site by forming the highly distorted octahedron and reduces the asymmetry around A site. The structure of Dy$_{2}$Zr$_{2}$O$_{7}$ is slightly equivalent to disordered fluorite with x = 0.3662 ($\Delta$x = 2.35$\%$). The x(O) position is directly related to the structure and strongly affect the $<$Zr-O-Zr$>$ and $<$Dy-O-Dy$>$ bond angle which are 72.41$^{o}$ and 109$^{o}$ respectively. In particular, $<$B-O-B$>$ angle for fluorite structure is 109$^{o}$ and for pyrochlore structure, it increases to 120 - 130$^{o}$. The obtained $<$Dy-O-Dy$>$ bond angle favor the formation of pyrochlore phase and $<$B-O-B$>$ bond angle is small but closer to the other members of pyrochlore family \cite{subramanian1983oxide, garbout2018pyrochlore}.

Further the stability of pyrochlore structure can be estimated by the ratio of cationic radii (r$_{A}$/r$_{B}$) \cite{yamamura2003electrical}. For pyrochlore structure, r$_{A}$/r$_{B}$ ranges between 1.48 to 1.78. The deviation whether smaller/larger than these values leads to the formation of defect-fluorite/perovskite type structure respectively \cite{pal2018high}. For Dy$_{2}$Zr$_{2}$O$_{7}$, the r$_{A}$/r$_{B}$ ratio of 1.43 (where r$_{A}$ = 1.027 $\AA$ and r$_{B}$ = 0.72 $\AA$ for Dy$^{3+}$ and Zr$^{4+}$ respectively) indicates the formation of defect fluorite structure. On the substitution of La$^{+3}$ in Dy$_{2}$Zr$_{2}$O$_{7}$, the r$_{A}$/r$_{B}$ ratio shifted towards pyrochlore regime (See table 1).

We have shown room temperature Raman spectra of Dy$_{2-x}$La$_{x}$Zr$_{2}$O$_{7}$ in figure 2. The Raman spectrum of fluorite is known to possess a single broad band (T$_{2g}$) because of the random distribution of oxygen ions over the anion sites \cite{hozoi2014longer}, whereas pyrochlore lattice exhibits six active Raman modes (A$_{1g}$, E$_{g}$, 4T$_{2g}$) \cite{han2015electron,hasegawa2010raman}. ForDy$_{2}$Zr$_{2}$O$_{7}$, Raman bands are observed at 123 cm$^{-1}$, 323 cm$^{-1}$, 401 cm$^{-1}$, 595 cm$^{-1}$ and 650 cm$^{-1}$. We can assign the Raman mode of Dy$_{2}$Zr$_{2}$O$_{7}$ at 323 cm$^{-1}$  (due to B-O$_{6}$ banding vibrations) to T$_{2g}$, 401 cm$^{-1}$ (mostly B-O, A-O stretching and O-B-O bending vibrations) to E$_{g}$ and 595 cm$^{-1}$ (mostly B-O stretching) to T$_{2g}$ \cite{vandenborre1983rare,sayed2011sm2}. The single broad band at $\sim$ 466 cm$^{-1}$ corresponding to the fluorite structure was not observed by us \cite{hozoi2014longer}. A relatively weak band observed at 123 cm$^{-1}$ can be associated with disorder in the system or anharmonic effect \cite{vandenborre1983rare} and band at 650 cm$^{-1}$ can be attributed to the higher coordination of B ion \cite{glerup2001structural}.

\begin{figure}[ht]
	\begin{center}
		\vspace{-10pt}
		\includegraphics[scale= 0.3]{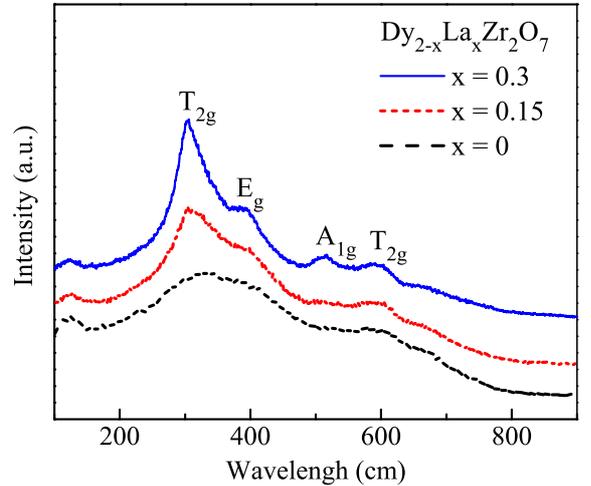}
		\vspace{-10pt}
		\caption{Room temperature Raman spectra of Dy$_{2-x}$La$_{x}$Zr$_{2}$O$_{7}$; x = 0, 0.15 and 0.3, showing the presence of Raman modes corresponding to pyrochlore structure.}
		\vspace{-25pt}
	\end{center}
\end{figure}

Thus, it can be inferred that although a substantial global disorder is seen from XRD measurements, ordering at microdomains seen from Raman spectra favors the weak pyrochlore structure in Dy$_{2}$Zr$_{2}$O$_{7}$. The observation of the remnants of pyrochlore modes in Raman data suggests it as a weak pyrochlore at the microdomain level and the defect fluorite structure in bulk. Based on X-Ray, Raman and synchrotron radiation XRD techniques Mandal \textit{et al.} has shown that Dy$_{2}$Hf$_{2}$O$_{7}$ is a weakly ordered pyrochlore which possess remnants of pyrochlore in Raman spectra and presence of superstructure peaks in the XRD \cite{mandal2006preparation}.\\
Upon substitution of La, all the five pyrochlore Raman modes become sharp and intense. Additionally, La substituted compounds shows one extra peak at 512 cm$^{-1}$ which is marked as A$_{1g}$. This mode has particular importance in pyrochlores as it is directly related to the trigonal distortion of B-O$_{6}$ octahedra through the modification of $<$B-O-B$>$ bond angle \cite{taniguchi2004raman}. These results suggest that the structural symmetry increases with substitution and leads the system towards weakly ordered pyrochlore phase.\\

\begin{figure}[htbp]
	\begin{center}
		\vspace{-20pt}
		\includegraphics[scale = 0.3]{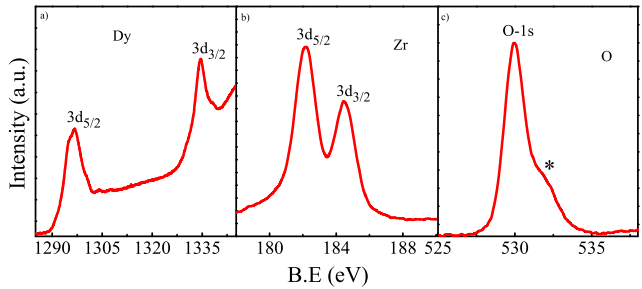}
		\vspace{-10pt}
		\caption{XPS spectra of Dy, Zr and O ions of Dy$_{2}$Zr$_{2}$O$_{7}$ collected at room temperature}
			\vspace{-25pt}
	\end{center}
\end{figure}

The x-ray photoemission spectroscopy (XPS) measurement on Dy$_{2}$Zr$_{2}$O$_{7}$ (figure 3) confirmed +3 and +4 oxidation state of Dy and Zr ion as that of Dy and Ti in Dy$_{2}$Ti$_{2}$O$_{7}$ \cite{snyder2001spin}. The xps spectra of oxygen contains two signals. The first signal (O-1s) at $\sim$ 529 eV corresponds to lattice oxygen and a very weak signal (marked by star) at $\sim$ 532 eV indicates defects/vacancies or surface chemi-absorbed oxygen \cite{tang2006mnox}. The ordered oxygen vacancies of the pyrochlores might be the possible reason for the observation of a weak signal at $\sim$ 532 eV.

\subsection{DC Magnetization}

DC magnetization of the compounds (figure 4) was measured in the temperature range T = 1.8 - 300 K in zero field cooled (ZFC) and field cooled (FC) protocol at the magnetic field (H) of 5 kOe. Similar to Dy$_{2}$Ti$_{2}$O$_{7}$, dc susceptibility $\chi_{dc}(T)$ of Dy$_{2-x}$La$_{x}$Zr$_{2}$O$_{7}$ increases monotonically on cooling and does not show any anomaly down to 1.8 K \cite{anand2015investigations}. Interestingly, $\chi_{dc}(T)$ shows lowering in the value below 15 K for the high fields of H = 10 kOe and 20 kOe (right inset of figure 4) and indicates the development of magnetic spin correlation for T $\leq$ 15 K. However, the absence of thermal hysteresis in ZFC and FC curves rules out the formation of magnetic cluster or spin glass like freezing at this temperature in these compounds.

\begin{figure}[ht]
	\begin{center}
		\vspace{-10pt}
		\includegraphics[scale=0.3]{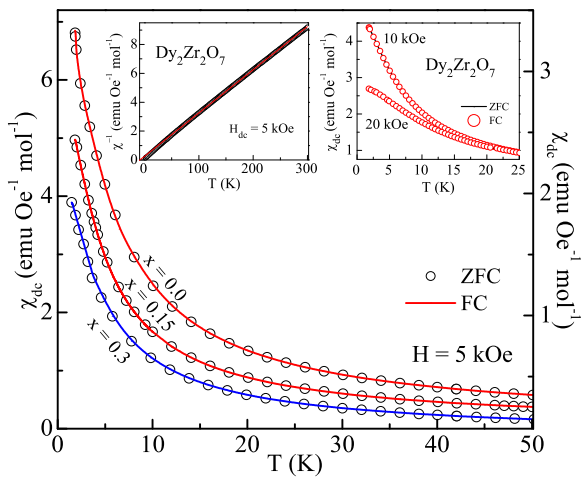}
		\vspace{-10pt}
		\caption{(Color line) DC magnetic susceptibility $\chi_{dc}$ versus T at H = 5 kOe for Dy$_{2-x}$La$_{x}$Zr$_{2}$O$_{7}$. Inset (Left): Curie-Weiss fitting (red line) of $\chi^{-1}$ (T) in 30 $\leq$ T $\leq$ 300 K at H = 5 kOe. Inset (Right): $\chi_{dc}$(T) versus T at H = 10 kOe and 20 kOe for Dy$_{2}$Zr$_{2}$O$_{7}$.}
	\end{center}
\vspace{-20pt}
\end{figure}

The Curie-Weiss $\chi$ = C/(T - $\theta_{CW}$) fit of $\chi^{-1}$(T) data for Dy$_{2}$Zr$_{2}$O$_{7}$ in the high temperature above 30 K (left inset of figure 4) gives $\theta_{CW}$ = -9.6(2) K, $\mu_{eff}$ = 16.41 $\mu_{B}$ and the same fitting in the low temperature range below 30 K yields $\theta_{CW}$ = -1.9(4) K.  With the inclusion of the correction for demagnetization factor for powdered compound, the readjusted $\theta_{CW}$ values are -8.1(2) K and -0.5(4) K for the mentioned temperature ranges, which indicate AFM interactions between Dy$^{3+}$ spins \cite{bramwell2000bulk}.  The obtained value of $\mu_{eff}$ $\sim$ 8.20 $\mu_{B}$/Dy for Dy$_{2}$Zr$_{2}$O$_{7}$ is significantly lower than the expected paramagnetic state value of $\mu_{eff}$ = g$_{J}\sqrt{J(J+1)}$ = 10.64 $\mu_{B}$/Dy per free Dy$^{3+}$ ions (g$_{J}$ = 4/3 and J = 15/2). The reduced value of $\mu_{eff}$ reflects the Ising anisotropic nature of magnetic ground state and has been attributed to the crystal-field effect \cite{anand2015investigations,anand2015observation}.

It is to note that the $\theta_{CW}$ = +0.5 K  and 1.9 K have been reported for Dy$_{2}$Ti$_{2}$O$_{7}$ and Ho$_{2}$Ti$_{2}$O$_{7}$ respectively from the similar temperature range (1.8 - 30 K) \cite{ramirez1999zero,anand2015investigations}. These compounds possess AFM exchange interaction at local Dy-Dy level but depict FM exchange interaction at global level with positive $\theta_{CW}$ \cite{ramirez1999zero,anand2015investigations}. It is worthy to mention that Pr$_{2}$Zr$_{2}$O$_{7}$ also shows negative $\theta_{CW}$ and owing to smaller value of $\mu_{eff}$ ($\sim$ 2.5 $\mu_{B}$/Pr) it is considered as the exchange spin ice material \cite{matsuhira2009spin}. The substitution of La for x = 0.15 and 0.3 in Dy$_{2}$Zr$_{2}$O$_{7}$, substantially reduces the $\mu_{eff}$ and $\theta_{CW}$ (See Table 1).

We further estimated the nearest neighbor dipole-dipole interactions D$_{nn}$ and exchange interactions J$_{nn}$ for Dy$_{2-x}$La$_{x}$Zr$_{2}$O$_{7}$. The dipole-dipole interactions can be estimated using the equation \cite{xu2015magnetic}
\begin{eqnarray}
D_{nn} = \frac{5}{3}\bigg(\frac{\mu_o}{4\pi}\bigg)\frac{\mu_{eff}^{2}}{r^{3}_{nn}}
\end{eqnarray}
Using the distance between nearest neighbor Dy$^{3+}$ ions, r$_{nn}$ = (a/4)/$\sqrt{2}$ = 3.70 $\AA$ (a = lattice constant) and $\mu_{eff}$ = 8.20 $\mu_{B}$/Dy, we obtained D$_{nn}$ = 1.38 K for Dy$_{2}$Zr$_{2}$O$_{7}$. To determine the nearest neighbor exchange interactions J$_{nn}$, the $\chi_{dc}$(T) data was fitted by $\chi_{dc}$(T) = (C$_{1}$/T)[1 + C$_{2}$/T], where C$_{1}$ is the curie constant and C$_{2}$ consists of the exchange and dipolar terms between the $<$111$>$ Ising moments (C$_{2}$ = (6S$^{2}$/4)[2.18D$_{nn}$ + 2.67J$_{nn}$]) in the linear low temperature region 10 - 30 K \cite{siddharthan1999ising}. The fitting yields C$_{2}$ = -1.55(03) and using D$_{nn}$ = 1.38 K, we obtained J$_{nn}$ = -1.188 K. The ratio J$_{nn}$/D$_{nn}$ $\sim$ -0.86 is close to the value for spin ice (-0.91), and thus favors the spin ice behavior \cite{den2000dipolar}. Byron C. \textit{et al.} has explained the occurrence of spin ice behavior in Dy$_{2}$Ti$_{2}$O$_{7}$ and Ho$_{2}$Ti$_{2}$O$_{7}$ despite the presence of long-range dipolar interactions and showed that the spin ice behavior is recovered over a large range of J$_{nn}$/D$_{nn}$ parameter \cite{den2000dipolar}. The J$_{nn}$/D$_{nn}$ ratio for these spin ice candidates are -0.53 and -0.22 respectively \cite{bramwell2001spin}. The J$_{nn}$/D$_{nn}$ ratio increases with La substitution (See Table 1) and is expected to stabilize the spin ice phase in the substituted compounds. 

In order to determine an accurate value of the ratio J$_{nn}$/D$_{nn}$, we have estimated the nearest neighbor Dy-Dy exchange interaction J$_{nn}$ from first principles DFT calculations for the parent compound Dy$_{2}$Zr$_{2}$O$_{7}$. J$_{nn}$ was calculated by mapping the total energy difference to a classical Heisenberg spin Hamiltonian \cite{xiang2011single}. We considered four different spin configurations of two neighboring Dy$^{3+}$ spins: (i) ${\uparrow}$ ${\uparrow}$ (ii) ${\downarrow}$ ${\downarrow}$ (iii) ${\uparrow}$ ${\downarrow}$ and (iv) ${\downarrow}$${\uparrow}$ with corresponding total energies given by E$_{1}$, E$_{2}$, E$_{3}$ and E$_{4}$ respectively keeping all other spins ferromagnetically aligned. The J$_{nn}$ is then calculated by following formula 
\begin{eqnarray}
	J_{nn} = ({E_{1}+E_{2}-E_{3}-E_{4}})/{4nS^{2}}
\end{eqnarray}
where S = 15/2, n = 2. The calculated near neighbor exchange interaction within GGA+SO approximation is found to be AFM in nature and the value is -1.276 K. For this value of J$_{nn}$, we obtain the ratio J$_{nn}$/D$_{nn}$ to be -0.92 which lies in the range of spin ice systems \cite{den2000dipolar}. Thus, our results substantiates the possibility of spin ice behavior for Dy$_{2}$Zr$_{2}$O$_{7}$. It is to mention that even without the application of Coulomb correlation $U$, we obtain magnetic moment value at Dy site to be 4.9$\mu_B$ with the value of J$_{nn}$/D$_{nn}$ = -0.92 for U$_{eff}$ = 0 being the closest to the spin ice value.

\begin{figure}[ht]
	\begin{center}
		\includegraphics[scale= 0.3]{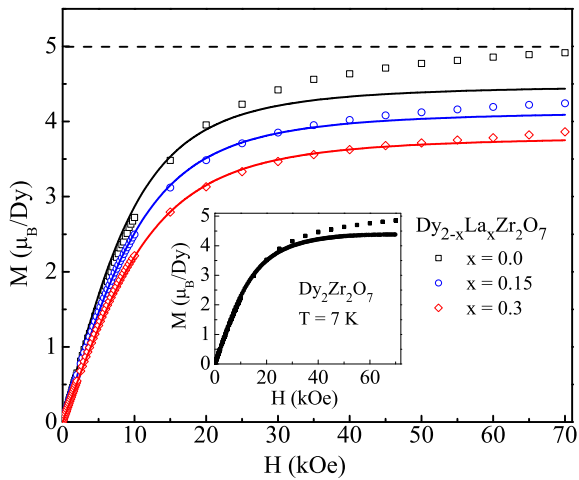}
		\vspace{-7pt}
		\caption{(Open symbol) Isothermal Magnetization M(H) of Dy$_{2-x}$La$_{x}$Zr$_{2}$O$_{7}$; x = 0, 0.15, 0.3 at 5 K. The solid curves are the fits of M(H) data by Eq. (3). Inset: Isothermal Magnetization M(H) of Dy$_{2}$Zr$_{2}$O$_{7}$ at T = 7 K along with fitting by Eq. (3) with an effective longitudinal g factor g$_{zz}$ = 18.4(02).}
		\vspace{-30pt}
	\end{center}
\end{figure}

The isothermal M(H) for Dy$_{2-x}$La$_{x}$Zr$_{2}$O$_{7}$ measured at T = 5 K are shown in Figure 5. Initially M increases rapidly with field and tends towards saturation above H $\geq$ 30 kOe in the substituted compounds only. The saturation magnetization (M$_{s}$) value of 4.80 $\mu_{B}$/Dy for Dy$_{2}$Zr$_{2}$O$_{7}$ at 70 kOe is close to  M$_{s}$ = 5 $\mu_{B}$/Dy for Dy$_{2}$Ti$_{2}$O$_{7}$ \cite{fukazawa2002magnetic}. These values are $\sim 45\%$ lower than the free ion theoretical value of M$_{s}$ = 10.64 $\mu_{B}$/Dy, and indicate the strong Ising anisotropy similar to a local $<$111$>$ Ising anisotropic system \cite{anand2015investigations, anand2015observation}. The M$_{s}$/Dy values decrease further for the La substituted Dy$_{2}$Zr$_{2}$O$_{7}$. The reduction in M$_{s}$/Dy ($\sim 4 -25\%$) in comparison to Dy$_{2}$Ti$_{2}$O$_{7}$ points to enhanced anisotropy in Dy$_{2-x}$La$_{x}$Zr$_{2}$O$_{7}$ compounds \cite{ fukazawa2002magnetic}. It is to note that M(H) value (at T = 5 K) reaches to saturation state value of $<$M$>$ = 3.33$\mu_{B}$/Dy, for the field of H $\approx$ 10 - 20 kOe only, which is required to comply with the ice rule along [111] axis. For higher fields, magnetization saturates (M$_{s}$ = 4.80 $\mu_{B}$/Dy) for Dy$_{2}$Zr$_{2}$O$_{7}$ in accordance with the three-in, one-out configuration g$_{J}$J{(1 + (1/3)$\times$3)/4} = 5 $\mu_{B}$/Dy \cite{anand2015investigations}. This large value of magnetization along [111] axis compared to the ice rule expected value (3.33 $\mu_{B}$/Dy) confirms the presence of strong Ising anisotropy in the system which results in the breakdown of ice rule \cite{anand2015investigations}. The magnetization does not reach to the saturating value up to 70 kOe in Dy$_{2}$Zr$_{2}$O$_{7}$, whereas in La substituted Dy$_{2}$Zr$_{2}$O$_{7}$ saturation is attained at lower field that indicate the reduction in disorder in the substituted compounds and favor the stabilization of pyrochlore phase on La substitution.

Considering the effective spin half doublet ground state system with local $<$111$>$ Ising anisotropy with transverse and longitudinal g-factors (g$_{\perp}$ = 0 and g$_{||}$ = g$_{zz}$), the average magnetization of system is given by
\begin{eqnarray}
<M>  = \frac{(k_{B}T)^{2}}{g_{zz}\mu_{B}H^{2}S}\int_{0}^{\frac{g_{zz}\mu_{B}HS}{k_{B}T}}x\tanh(x)dx
\end{eqnarray}
where x = g$_zz$$\mu_{B}$HS/k$_{B}$T. The M(H) data shows a good fit with above expression for Dy$_{2}$Zr$_{2}$O$_{7}$ at lower field only. The fitting improves for higher temperature isotherms (See inset of Figure 5) and La substituted compounds. The obtained value of g$_{zz}$ = 18.05(02) for Dy$_{2}$Zr$_{2}$O$_{7}$ at 5 K is close to the g$_{zz}$ = 2g$_{J}$J = 20, as expected for the pure Kramers doublet m$_{J}$ = $\pm$15/2 states of Dy$^{3+}$. A closer value of g$_{zz}$ = 18.5(1) is reported for Dy$_{2}$Ti$_{2}$O$_{7}$ in Ref\cite{bramwell2000bulk}. The value of the ground state moment of Dy $\sim$ 9.01(2) $\mu_{B}$/Dy for Dy$_{2}$Zr$_{2}$O$_{7}$ (calculated using m$_{Dy}$ = g$_{zz}$S$\mu_{B}$) is consistent with the $\mu_{eff}$ determined from the dc magnetization. The g$_{zz}$ value increases from $\sim$ 16 at 1.8 K to $\sim$ 18.4 at 7 K for Dy$_{2}$Zr$_{2}$O$_{7}$, and indicates the admixture of additional terms in the m$_{J}$. The lowering of g$_{zz}$ in La substituted compounds further points towards the enhanced anisotropy. The poor fitting of the M(H) data at low temperature and high field using the spin 1/2 Ising model highlights the need for additional anisotropic terms and non-zero transverse g-factor in the model for these compounds. These data indicate that Dy$_{2}$Zr$_{2}$O$_{7}$ has dominant AFM coupling between nearest-neighbor Dy ions and easy axis anisotropy.

\subsection{AC susceptibility}

\begin{figure}[ht]
	\begin{center}
		\includegraphics[scale=0.3]{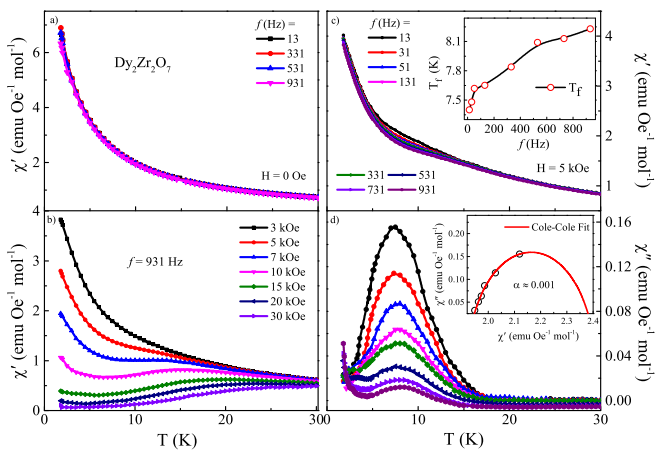}
		\vspace{-20pt}
		\caption{(a) The $\chi^{\prime}$(T) of Dy$_{2}$Zr$_{2}$O$_{7}$ measured at H = 0 for frequencies \textit{f} = 10 Hz to 1 kHz. (b) T dependence of $\chi^{\prime}$ at $\it{f}$ = 931 Hz measured at different fields between H = 3 kOe to 30 kOe. Figure 6c and 6d shows the $\chi^{\prime}$(T) and $\chi^{\prime\prime}$(T) of Dy$_{2}$Zr$_{2}$O$_{7}$ measured at different frequencies between 10 Hz - 1 kHz and H = 5 kOe. Inset of 6c and 6d are the frequency dependence of freezing temperature and Cole-Cole fit (at T = 7.5 K) respectively.}
			\vspace{-20pt}
	\end{center}
\end{figure}

Figure 6a shows the temperature dependence of ac susceptibility $\chi_{ac}$(T) of Dy$_{2}$Zr$_{2}$O$_{7}$ measured at different frequencies between 10 - 1000 Hz at zero applied dc field. The real part of ac susceptibility ($\chi^{\prime}$) of Dy$_{2}$Zr$_{2}$O$_{7}$ (figure 6a) measured in zero dc magnetic field  does not show any anomaly down to 1.8 K. The large value of $\chi^{\prime}$ suggests the liquid like dynamic state of the system at low temperature which has been studied in detail by Ramon \textit{et al.} \cite{ramon2019absence}. We measured $\chi_{ac}$ in the presence of dc field also. The $\chi^{\prime}$(T) measured at $\it{f}$ = 931 Hz at H = 3 - 30 kOe is shown in figure 6b. The $\chi^{\prime}$(T) is similar to zero field data, up to 3 kOe field, however a weak anomaly starts developing near T $\approx$ 10 K for H = 5 kOe field. The deviation in $\chi^{\prime}$ from paramagnetic-like behavior leads to a sharp rise in $\chi^{\prime\prime}$ at the freezing temperature, which increases with frequency. For the further increase in field, the anomaly becomes more prominent and is pushed towards higher temperature with significant decrease in magnitude above 10 kOe. The slowing down of spin dynamics on the application of field for T $<$ T$_{f}$ reveals magnetic spin freezing similar to the spin ice compounds: Dy$_{2}$Ti$_{2}$O$_{7}$ and Ho$_{2}$Ti$_{2}$O$_{7}$ \cite{snyder2002dirty, matsuhira2001novel, harris1997geometrical}. A closer value of T$_{f}$ to the $\theta_{CW}$ obtained from dc magnetization data, suggesting the development of weak AFM interactions for high magnetic field.

In order to examine the nature of spin freezing transition, we performed the $\chi_{ac}$(T) measurement at H = 5 kOe for different frequencies between 10 Hz to 1000 Hz (shown in figure 6c and 6d). As seen from the figures, a clear anomaly is present in $\chi^{\prime}$(T) and $\chi^{\prime\prime}$(T) below $\sim$15 K at 13 Hz frequency. For the further increase in frequency, maximum in $\chi^{\prime\prime}$(T) shift towards higher temperature. This feature is a signature of glassy transition in systems, having competing interactions or structural defects. To analyze the glass-like freezing of the spin degree of freedom, we have quantified the frequency dependence using the relation $\it{p}$ = $\Delta$T$_{f}$/T$_{f}$$\Delta$(log$\it{f_{o}}$). The obtained value of $\it{p}$ $\approx$ 0.08, is much higher than the typical spin glass ($\it{p}$ $\approx$ 0.01) \cite{snyder2001spin}. The rise in freezing temperature with field also rule out the possibility of spin glass like freezing, where the application of magnetic field suppress the freezing temperature in spin glass.

The more striking difference between the glassy freezing and spin-ice freezing is found in the distribution of relaxation times. We have performed Cole-Cole analysis of $\chi^{\prime\prime}$ and $\chi^{\prime}$ data as shown in the inset of figure 6d. The Cole-Cole plot gives the distribution of relaxation times of spin-freezing process and depicts the spin dynamics at a given temperature. The $\chi^{\prime\prime}$ and  $\chi^{\prime}$ data is fitted within the Cole-Cole formalism, given by equation\cite{topping2018ac}
\begin{eqnarray}
\chi^{\prime\prime}(\chi^{\prime}) = -\frac{\chi_{o}-\chi_{s}}{2\tan\big[(1-\alpha)\frac{\pi}{2}]}\nonumber \\
+  \sqrt{(\chi^{\prime}-\chi_{s})(\chi_{o}-\chi^{\prime}) + \frac{(\chi_{o}-\chi_{s})^{2}}{4\tan^{2}\big[(1-\alpha)\frac{\pi}{2}]}}
\end{eqnarray} 

Here $\chi_{o}$ and $\chi_{s}$ are the isothermal and adiabatic susceptibility respectively. The parameter $\alpha$ depicts the width of the distribution of relaxation times, for single spin relaxation $\alpha$ = 0. From the best fit of $\chi^{\prime\prime}$ using equation 4, we obtained (1 - $\alpha$)$\pi$/2 = 89.88(1)$^{o}$. Accordingly, we obtain $\alpha$ $\approx$ 0.001, a very close value of the theoretically expected semicircle. The extremely low value of $\alpha$ and semi-circular character of the plot similar to Dy$_{2}$Ti$_{2}$O$_{7}$ and Ho$_{2}$Ti$_{2}$O$_{7}$ indicating the single spin relaxation or narrow distribution of relaxation time \cite{snyder2002dirty, matsuhira2001novel, harris1997geometrical}. This is unlike the magnetic systems exhibiting glass-like behavior, where the relaxation time is typically of several orders \cite{dekker1989activated,dirkmaat1987frequency,huser1983dynamical}. 

\begin{figure}[ht]
	\begin{center}
		\vspace{-10pt}
		\includegraphics[scale=0.3]{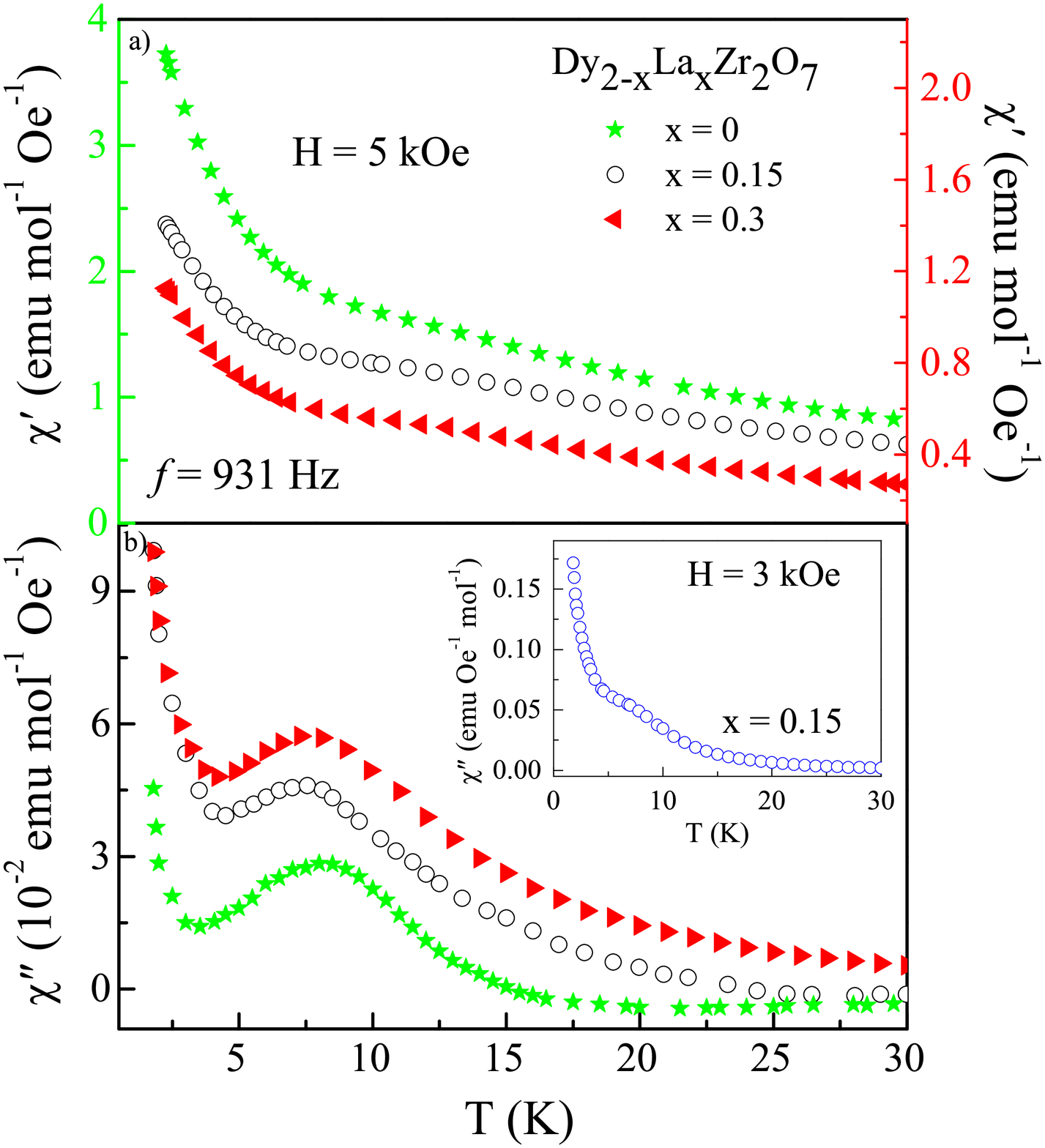}
		\vspace{-10pt}
		\caption{$\chi^{\prime}$(T) and $\chi^{\prime\prime}$(T) of Dy$_{2-x}$La$_{x}$Zr$_{2}$O$_{7}$; x = 0.0, 0.15, 0.3 measured at \textit{f} = 931 Hz and H = 5 kOe. Inset of 7b is the $\chi^{\prime\prime}$(T) of Dy$_{1.85}$La$_{0.15}$Zr$_{2}$O$_{7}$ at H = 3 kOe.}
	\end{center}
	\vspace{-15pt}
\end{figure}

Figure 7a and 7b show the real and imaginary part of $\chi_{ac}$(T) of  Dy$_{2-x}$La$_{x}$Zr$_{2}$O$_{7}$ for x = 0.0, 0.15 and 0.3 measured at H = 5 kOe. Similar to parent compound Dy$_{2}$Zr$_{2}$O$_{7}$, spin freezing anomaly is clearly visible in substituted compounds. These results are in agreement with the Raman spectroscopy studies insisting the strengthening of pyrochlore phase upon La substitution. As shown in the inset of figure 7b, the spin freezing anomaly develops at lower field H = 3 kOe for Dy$_{1.85}$La$_{0.15}$Zr$_{2}$O$_{7}$ compared to Dy$_{2}$Zr$_{2}$O$_{7}$ (H = 5 kOe). Similar studies on Ca or Y substituted Dy$_{2}$Ti$_{2}$O$_{7}$ show suppression in the freezing transition \cite{snyder2002dirty,anand2015investigations}. However, in the present case, non-magnetic La substitution stabilizes the pyrochlore phase and thus the spin ice state in the weakly ordered pyrochlore Dy$_{2-x}$La$_{x}$Zr$_{2}$O$_{7}$ system. Further the frequency dependence of ac susceptibility is not seen for Dy$_{2}$Zr$_{2}$O$_{7}$ at higher field (7 kOe). Although there is slight increase in $\chi^{\prime}$ value with increase in frequency near this anomaly, we do not observe any frequency dependence, ruling out the possibility of magnetic clustering, consistent with dc magnetization studies. (See supplementary information figure S3).

\subsection{Heat Capacity}

Figure 8 displays the magnetic contribution of the heat capacity (C$_{mag}$) of Dy$_{2}$Zr$_{2}$O$_{7}$ for H = 0 - 50 kOe. The C$_{mag}$(T) was calculated from total heat capacity C$_{p}$(T) after subtracting the lattice heat capacity. For the determination of lattice heat capacity, C$_{p}$(T) was fitted using Debye and Einstein models (see supplementary information figure S4). The C$_{mag}$(T) plotted on logarithmic T scale shows a peak at 1.2 K at H = 0. We have shown the data of Ramon \textit{et al.} for the comparison \cite{ramon2019absence}. Low temperature C$_{p}$(T) at H = 0 and 50 kOe field is shown in left inset of figure 8. As seen from  the figure 8, the C$_{mag}$ peak shifts towards higher temperature on the application of magnetic field and a similar behavior has been reported for Dy$_{2}$Ti$_{2}$O$_{7}$, Pr$_{2}$Zr$_{2}$O$_{7}$, Yb$_{2}$Ti$_{2}$O$_{7}$ etc.\cite{kadowaki2009observation,higashinaka2004low,petit2016antiferroquadrupolar,tachibana2007heat,thompson2017quasiparticle,hiroi2003specific}. It is to mention that Pr$_{2}$Zr$_{2}$O$_{7}$ does not exhibit any signature of long-range ordering in heat capacity and magnetic susceptibility measurements but still shows the evidence of spin correlation and quantum fluctuations \cite{petit2016antiferroquadrupolar,kimura2013quantum}. These results are interpreted as the indication of ice rule satisfied within the time scale of the measurements.
 
\begin{figure}[ht]
\begin{center}
\includegraphics[scale=0.3]{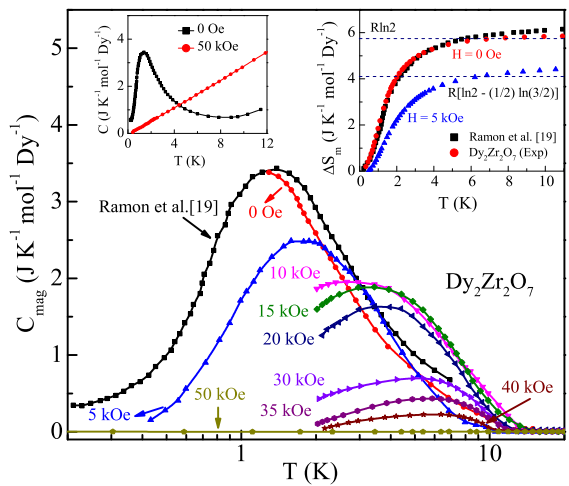}
\caption{ The C$_{mag}$(T) of Dy$_{2}$Zr$_{2}$O$_{7}$ for 0.4 $\leq$ T $\leq$ 30 K measured at 0 - 50 kOe. The C$_{mag}$(T) data from Ref\cite{ramon2019absence} is also shown for the comparison. Left inset shows the total heat capacity for Dy$_{2}$Zr$_{2}$O$_{7}$ at H = 0, 50 kOe. Right inset shows the recovered magnetic entropy as a function of temperature for Dy$_{2}$Zr$_{2}$O$_{7}$ at H = 0, 5 kOe. The dashed line denotes the expected entropy value for Ising spins and spin ice systems.}
\end{center}
\vspace{-10pt}
\end{figure}

\begin{figure}[ht]
\begin{center}
\includegraphics[scale= 0.3]{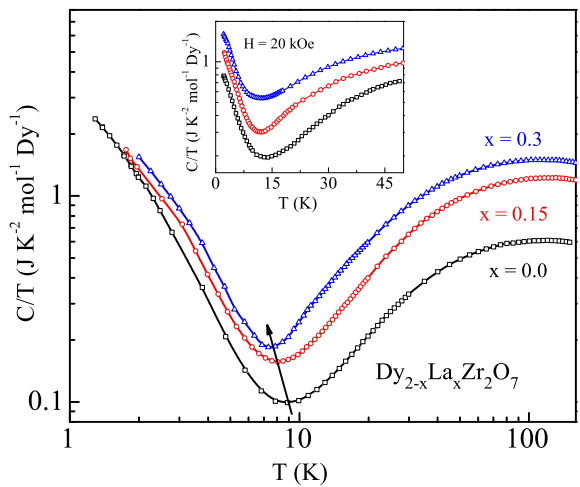}
\caption{(Color online) The C/T versus T measured at H = 0 Oe for Dy$_{2-x}$La$_{x}$Zr$_{2}$O$_{7}$; (x = 0.0, 0.15, 0.3). Inset shows the same for the compounds at H = 20 kOe.} 
\end{center}
\end{figure}
 
The magnetic entropy ($\Delta$S$_{m}$) of Dy$_{2}$Zr$_{2}$O$_{7}$ for H = 0 and 5 kOe is shown in the right inset of figure 8. The low temperature $\Delta$ S$_{m}$ value of Dy$_{2}$Zr$_{2}$O$_{7}$ at zero field is close to Rln2 (5.76 J/mole-K) which corresponds to the molar entropy of the classical system of two spin orientations per degree of freedom. The $\Delta$S$_{m}$ value decreases for higher fields and shows a remnant value of $\sim$ Rln2 - (1/2)Rln(3/2) (4.07 J/mole-K) for 5 kOe field, which is equivalent to the entropy of water ice \cite{giauque1933molecular,pauling1935structure} and other spin ice systems Dy$_{2}$Ti$_{2}$O$_{7}$, Ho$_{2}$Ti$_{2}$O$_{7}$ \cite{bramwell2001spin,ramirez1999zero}. This is quite remarkable as the slowing down of the spin dynamics under external magnetic field leads to the spin ice state. For higher field $\Delta$S$_{m}$ decreases further and almost no magnetic entropy is left for H $\ge$ 50 kOe. These results are consistent with the field induced spin freezing transition observed in ac susceptibility data at H = 5 kOe.  

The dilution of Dy$_{2}$Zr$_{2}$O$_{7}$ with non-magnetic La shows no significant change in heat capacity behavior except the shift of the spin freezing transition towards low temperatures (see figure 9). It is reported that the substitution of 10$\%$ Ca$^{2+}$ for Dy$^{3+}$ in Dy$_{2}$Ti$_{2}$O$_{7}$  does not change the spin freezing transition \cite{anand2015investigations}. Furthermore, similar to field dependence in Dy$_{2}$Zr$_{2}$O$_{7}$, the application of dc magnetic field (shown in the inset of figure 9 for H = 20 kOe) shifts the transition towards higher temperature for the La substituted compounds. Thus, the dilution of magnetism in Dy$_{2}$Zr$_{2}$O$_{7}$ by La substitution suppress the freezing transition whereas the dc magnetic field shifts the transition towards higher temperature. 

\section{Conclusion}

Our study explores the possibility of the spin ice phase in the highly disordered fluorite/pyrochlore compound Dy$_{2}$Zr$_{2}$O$_{7}$. The field dependent ac susceptibility studies suggest that the Dy$_{2}$Ti$_{2}$O$_{7}$ like spin ice phase can be induced in Dy$_{2}$Zr$_{2}$O$_{7}$ by quenching the structural disorder under well controlled magnetic field. The ac susceptibility measurements in dc applied field show the emergence of magnetic correlation/freezing near T $\sim$ 10 K which is akin to that observed for Dy$_{2}$Ti$_{2}$O$_{7}$. Semi-circular behavior of Cole-Cole plot and a reduced value of $\alpha$ suggests the presence of single spin relaxation. Further, the ratio of antiferromagnetic interactions and dipole-dipole ferromagnetic interaction J$_{nn}$/D$_{nn}$ $\sim$ -0.86 favors the spin ice state in the compound. As noticed in Ref\cite{higashinaka2005field} for Dy$_{2}$Ti$_{2}$O$_{7}$, the field induced magnetic transition is governed by the long-range dipolar interactions over the nearest neighbors. Similarly, in case of Dy$_{2}$Zr$_{2}$O$_{7}$, the J$_{eff}$ = J$_{nn}$ + D$_{nn}$ is large enough to favor the dominance of magnetic dipole-dipole interaction term. The smaller value of magnetic moment $\sim$ 8.2 $\mu_{B}$/Dy in comparison to expected 10.64 $\mu_{B}$/Dy; and saturation magnetization $\sim$ 4.8 $\mu_{B}$/Dy suggests strong Ising anisotropic nature of magnetic ground state.\\
The La substitution into Dy$_{2}$Zr$_{2}$O$_{7}$ stabilizes the weakly ordered pyrochlore structure, and strengthens the spin ice phase at even lower fields than required for Dy$_{2}$Zr$_{2}$O$_{7}$. Heat capacity results shows zero residual entropy at H = 0 Oe and presence of residual entropy of R/2ln(3/2) at H = 5 kOe indicating the stabilization of the spin ice state. On application of high magnetic field, the magnetic entropy of the system evolves from spin liquid state to water ice and then back to the non-magnetic ground state. It would be quite interesting to perform the field dependent in-elastic neutron scattering study to understand the dynamics of spin to ascertain the true magnetic state of these compounds.    
 
\section{Acknowledgment} The authors acknowledge Advanced Material Research Center (AMRC), IIT Mandi for the experimental facilities. Sheetal acknowledges IIT Mandi and Ministry of Human Resource Development (MHRD) India for Senior Research fellowship. We acknowledge the liquid Helium facility at IISER Mohali.

\bibliography{DZOsheetal}

\section{Supplementary Information}

\subsection{Crystal Structure}
\vspace{10pt}
\begin{table}[ht]
	\caption{Structural parameters of various atoms in Dy$_{2-x}$La$_{x}$Zr$_{2}$O$_{7}$ (x = 0, 0.15 and 0.3)}
	
	\begin{tabular}{cccccccccccccc}
		\hline
		Atom  &&&& site && x  & y  &&&& z && a \\
		\hline
		&&&&&&& Dy$_{2}$Zr$_{2}$O$_{7}$ Fd$\bar{3}$m && \\
		Dy  &&&& 16d && 0.5000 &  0.5000 &&&& 0.5000 &\\
		Zr  &&&& 16c && 0.0000 &  0.0000 &&&& 0.0000 && 10.4511(3)\\
		O$^{\prime}$  &&&& 48f && 0.3667(1) &  0.1250 &&&& 0.1250 &\\
		O  &&&& 8b && 0.3750  &  0.3750  &  &&& 0.3750 &\\
		&&&&&&& Dy$_{1.85}$La$_{0.15}$Zr$_{2}$O$_{7}$  && \\
		Dy  &&&& 16d && 0.5000 &  0.5000 &&&& 0.5000 &\\
		Zr  &&&& 16c && 0.0000 &  0.0000 &&&& 0.0000 && 10.4832(1)\\
		O$^{\prime}$  &&&& 48f  && 0.3674(0) &   0.1250 &&   && 0.1250 &\\
		O  &&&& 8b && 0.3750  &  0.3750  &  &&& 0.3750 &\\
		&&&&&&& Dy$_{1.7}$La$_{0.3}$Zr$_{2}$O$_{7}$  && \\
		Dy  &&&& 16d && 0.5000 &  0.5000 & &&& 0.5000 &\\
		Zr  &&&& 16c && 0.0000 &  0.0000 & &&& 0.0000 && 10.4971(2)\\
		O$^{\prime}$  &&&& 48f && 0.3692(2) &  0.1250 && && 0.1250 &\\
		O  &&&& 8b && 0.3750  &  0.3750  & &&& 0.3750 &\\
		&&&&&&&  Dy$_{2}$Zr$_{2}$O$_{7}$ Fm$\bar{3}$m  & &\\
		Dy &&&& 4a && 0.0000 & 0.0000 &&&& 0.0000 &\\
		Zr &&&& 4a && 0.0000 & 0.0000 &&&& 0.0000 && 5.228(3)\\
		O &&&& 8c && 0.2500 & 0.2500 &&&& 0.2500 &\\
		\hline
	\end{tabular}
\end{table}

\begin{figure}[ht]
	\begin{center}
		\includegraphics[scale=0.5]{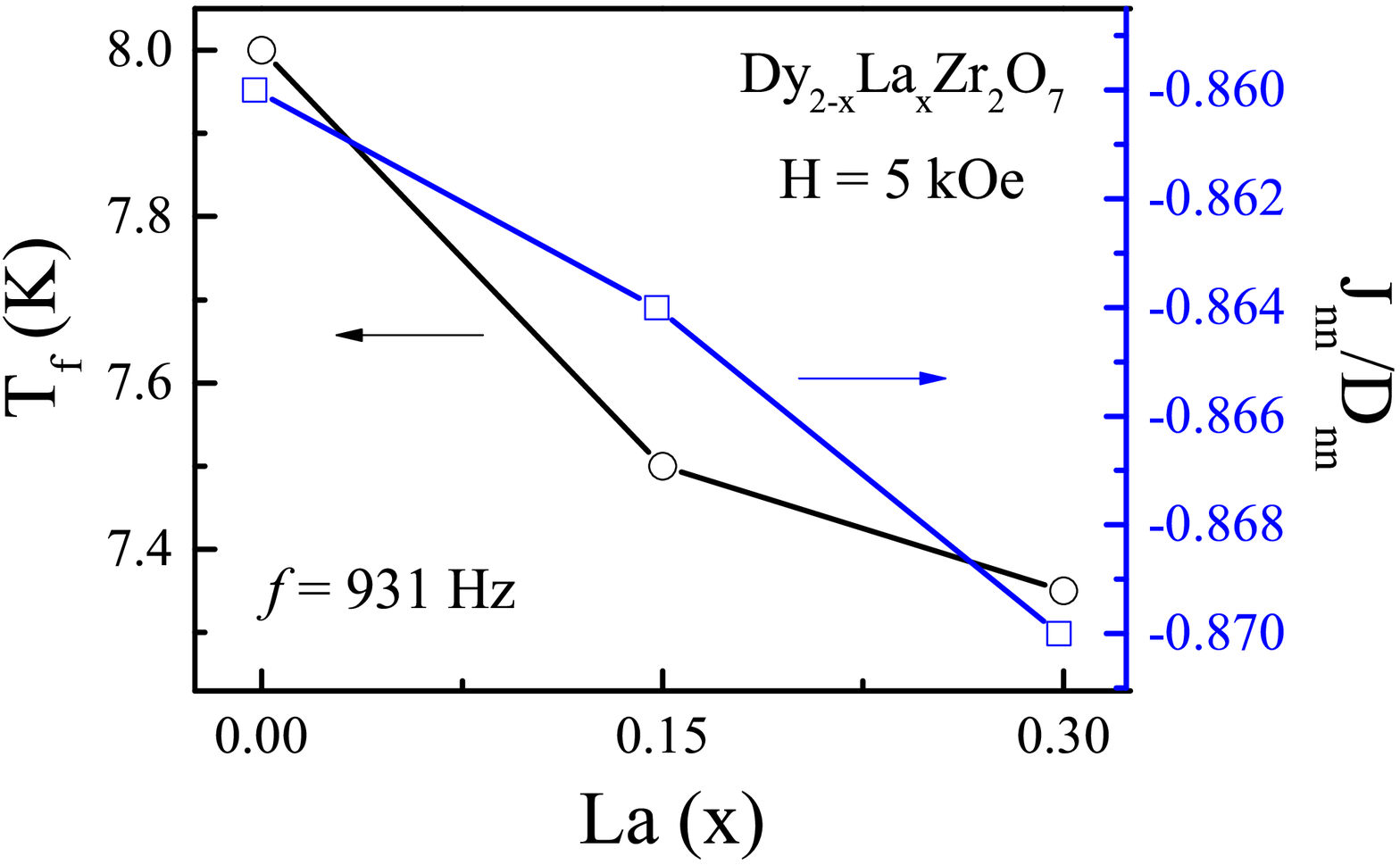}
		\vspace{1pt}
	\end{center}
\textbf{Figure S1}: Variation in freezing temperature (T$_{f}$) and J$_{nn}$/D$_{nn}$ ratio with La(x) concentration at frequency $\it{f}$ = 931 Hz and at a dc field of 5 kOe. It was observed that the freezing temperature suppressed slightly with increase in La concentration similar to the non-magnetic Ca and Y substitution on Dy site in Dy$_{2}$Ti$_{2}$O$_{7}$ \cite{snyder2002dirty, anand2015investigations}. The ratio of J$_{nn}$/D$_{nn}$ is remain in the spin-ice regime for all the substituted compounds and gets more closer to the optimum value $\sim$ -0.91 for the spin-ice systems \cite{den2000dipolar}.
\end{figure}

\newpage

\subsection{Dc susceptibility}

\begin{figure}[ht]
	\begin{center}
		\includegraphics[scale=0.35]{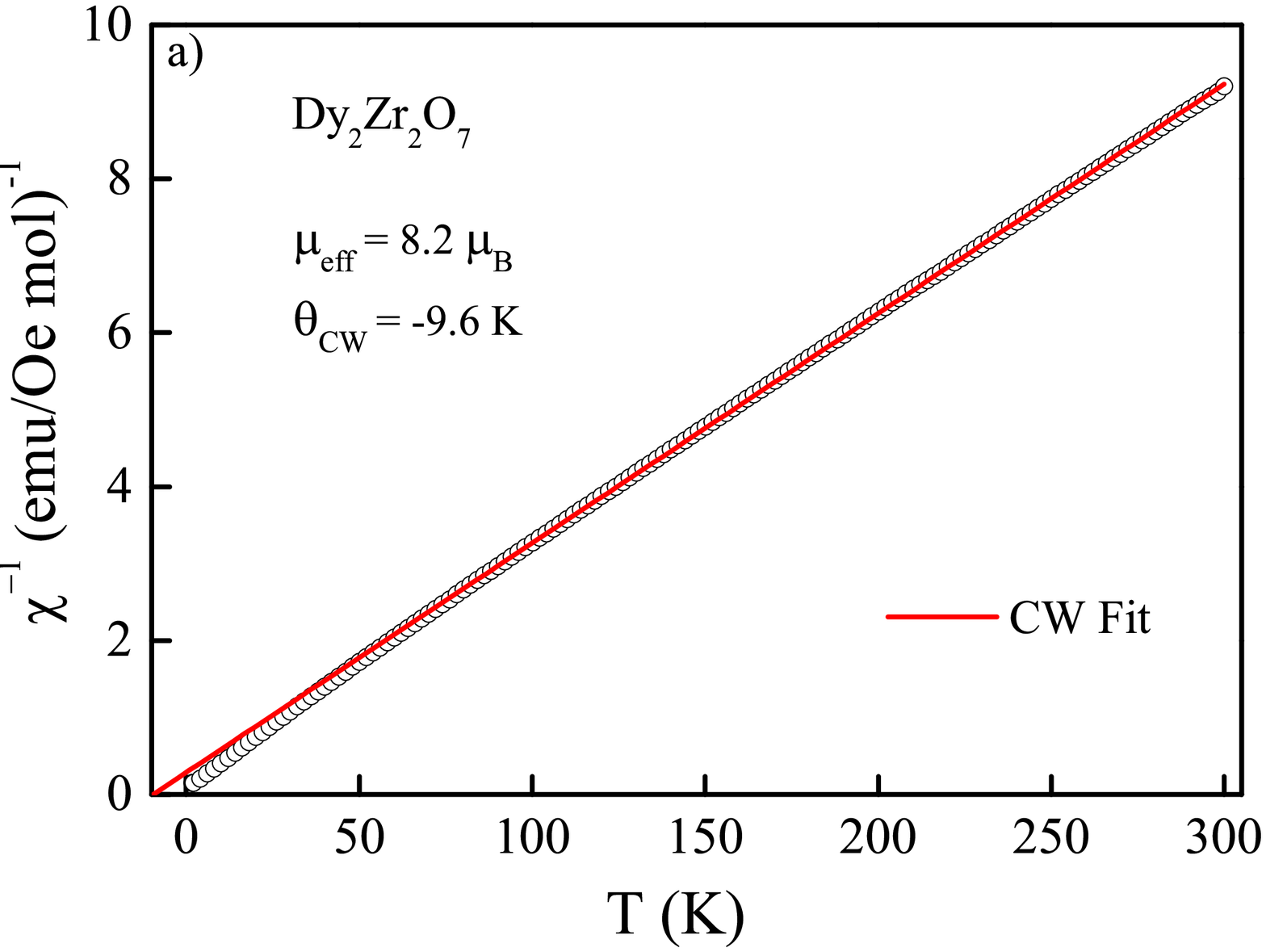}
		\vspace{-10pt}
		\includegraphics[scale=0.35]{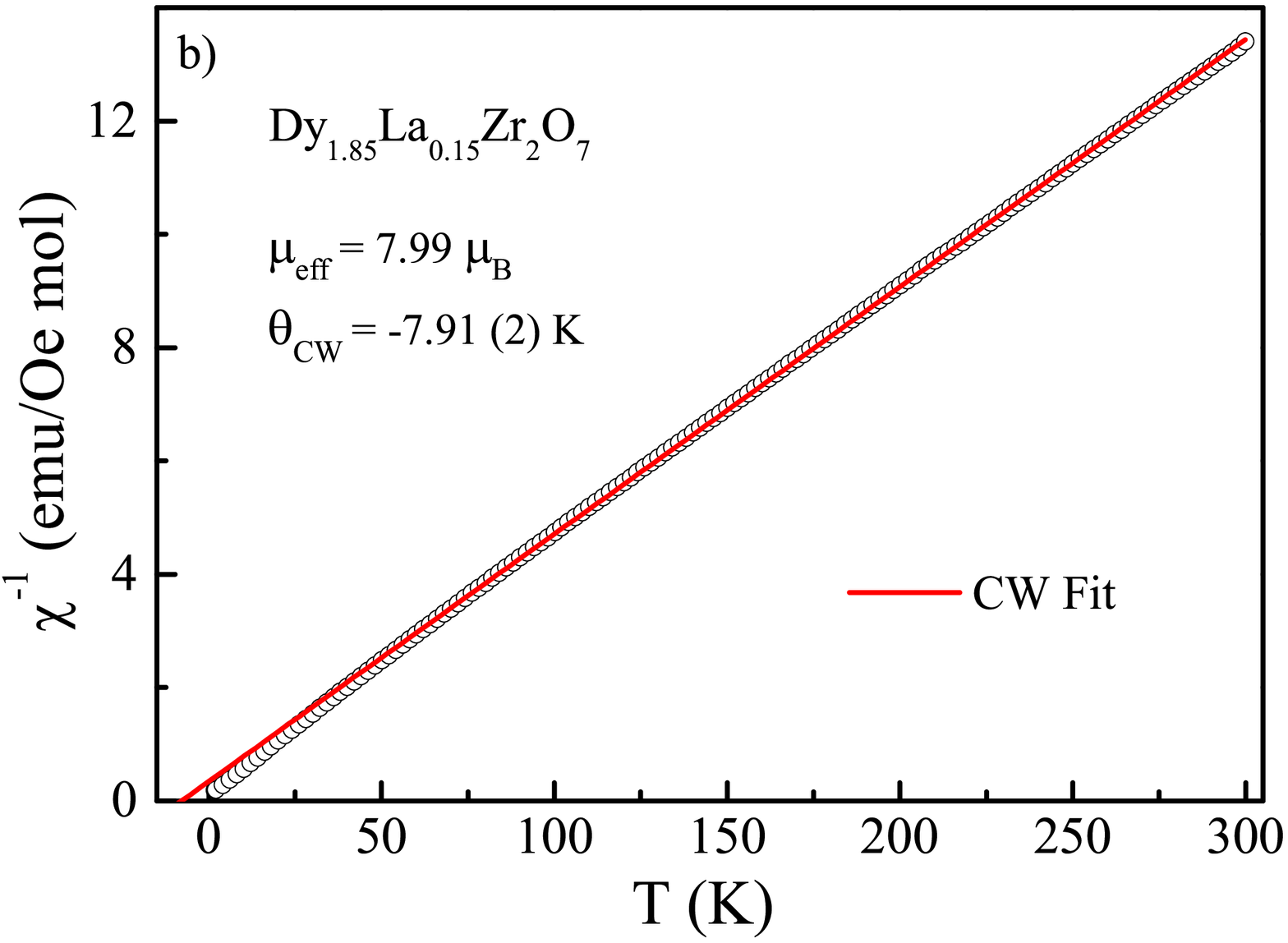}
		\vspace{-5pt}
			\includegraphics[scale=0.35]{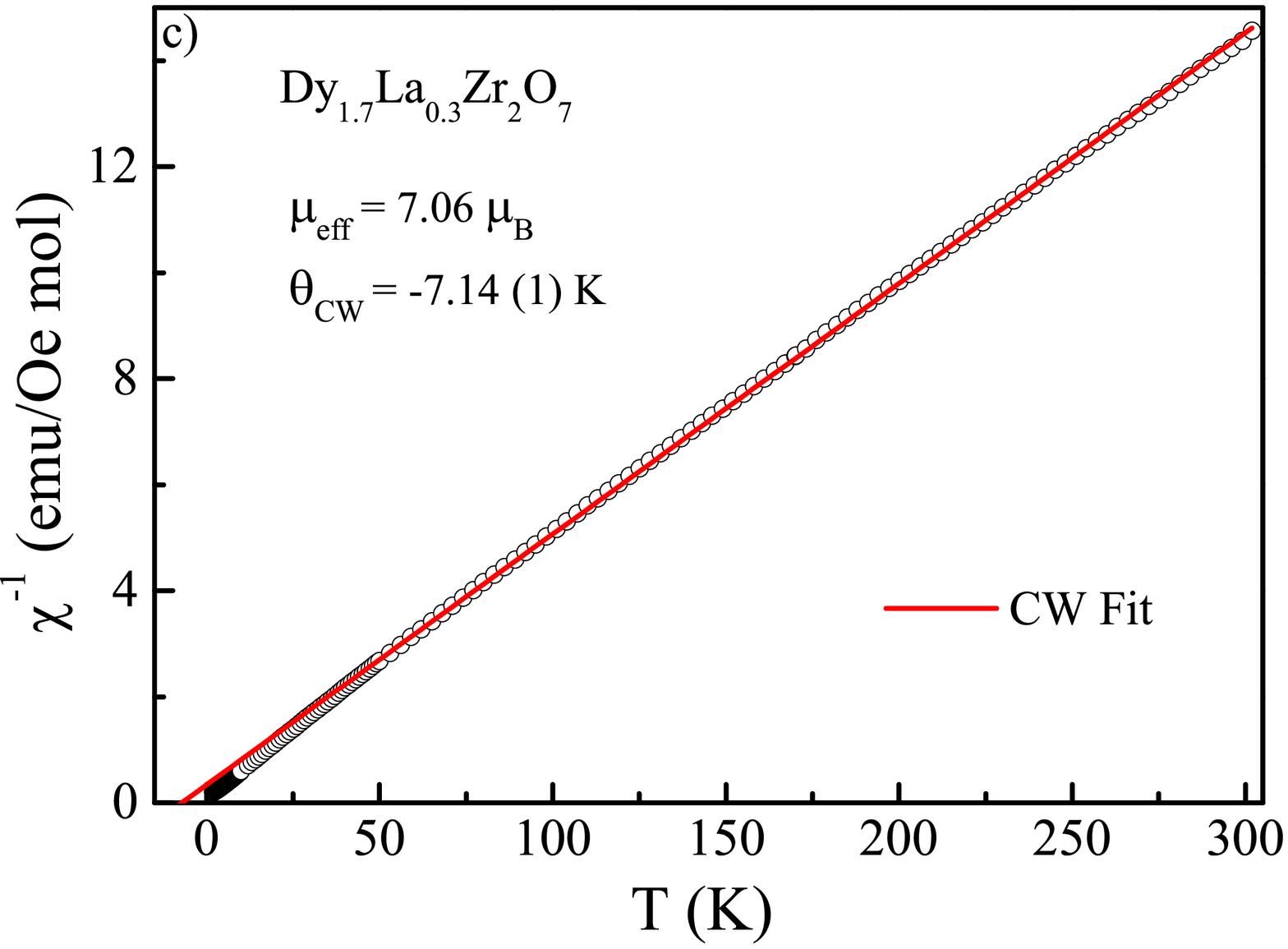}
		\vspace{-1pt}
	\end{center}

\textbf{Figure S3:} Curie-Weiss fit of dc magnetization data in the temperature range 30 - 300 K at H = 5 kOe for  Dy$_{2-x}$La$_{x}$Zr$_{2}$O$_{7}$ (x = 0, 0.15 and 0.3).
\end{figure}

\newpage

\subsection{Ac susceptibility}

\begin{figure}[ht]
	\begin{center}
		\vspace{-10pt}
		\includegraphics[width=7cm,height=12cm]{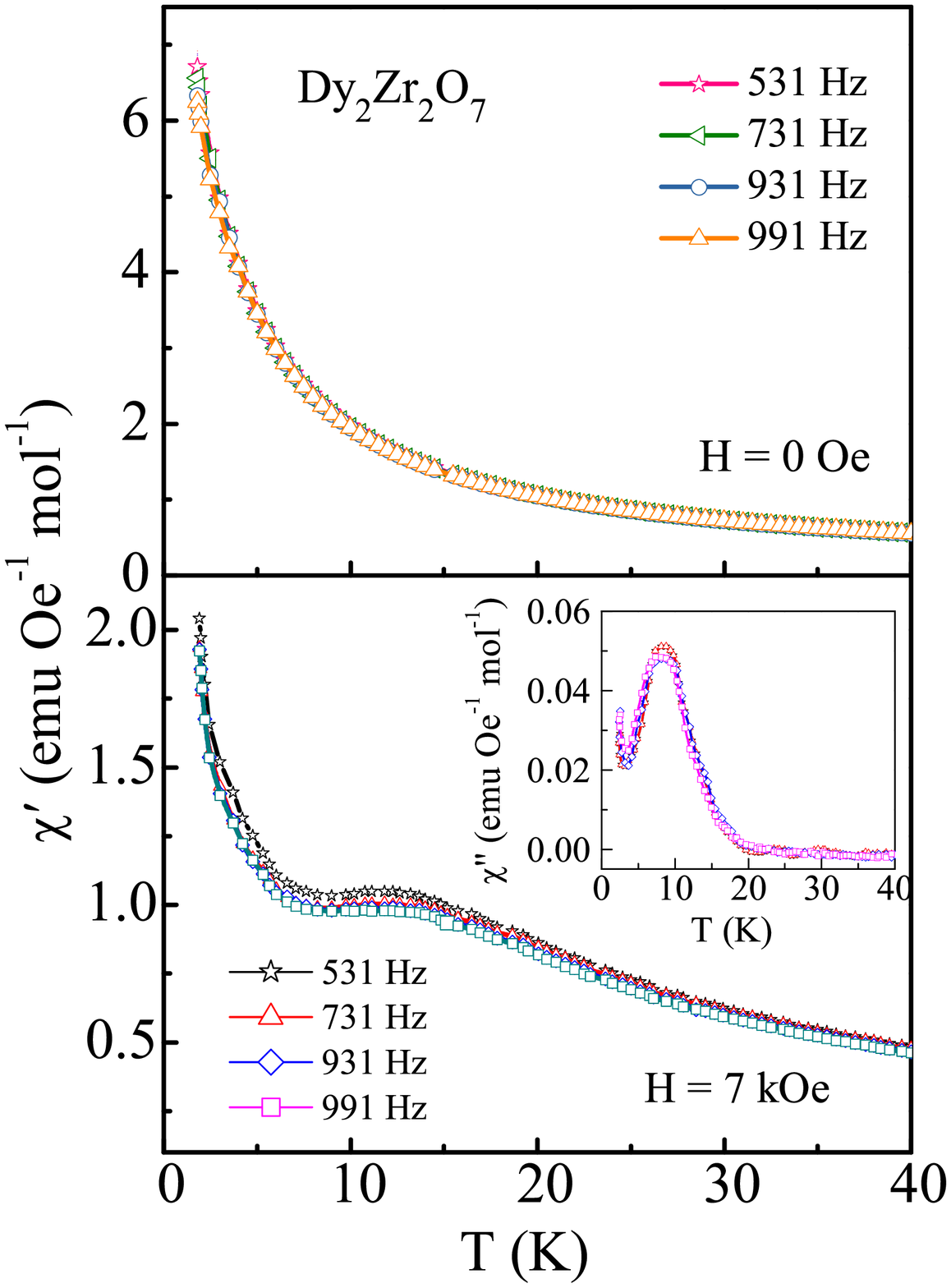}
		\vspace{-10pt}
		\end{center}
	\textbf{Figure S3:} (a) Temperature dependence of ac susceptibility $\chi^{\prime}$(T) of Dy$_{2}$Zr$_{2}$O$_{7}$ at zero dc magnetic field for \textit{f} = 531 - 931 Hz. (b) The $\chi^{\prime}$(T) of DZO measured at H = 7 kOe for frequencies \textit{f} = 531 - 931 Hz. Inset: The $\chi^{\prime\prime}$(T) of DZO measured at H = 7 kOe for frequencies \textit{f} = 531 - 931 Hz. 
\end{figure}

Real part of ac susceptibility increases monotonically with decreasing temperature without any observation of magnetic anomaly. As we superimposed dc magnetic field a field induced transition (H = 5 kOe) start developing at low temperature. We have done the complete anaylsis of this transition at 5 kOe in the main text. And to further study the nature of this field induced transition ac susceptibility measurement was performed at various frequencies in the presence of magnetic field of 7 kOe. On increasing the field the peak shifted to higher temperature but no significant change in peak position is observed with the increase in frequency  however, a slight change in amplitude occur. It is quite intersting and in support with the heat capacity data where the spin-ice entropy decreases with the application of higher magnetic field (H $\geq$ 5 kOe). The frequency independent behavior ruled out the possibility of any spin glass like transition\cite{snyder2001spin}. The anomaly  may indicates a different type of spin freezing as the pyrochlores have only the distorted spin geometry and negligible structural or chemical disorder ($<$1$\%$) which introduces the spin-glass behavior. However, to the best of our knowledge, there have been no clear explanation on the origin and nature of the lowest-temperature anomaly reported for other polycrystalline spin-ice compound Dy$_{2}$Ti$_{2}$O$_{7}$ at 16 K\cite{snyder2001spin,fukazawa2002magnetic,kadowaki2009observation}.

	\vspace{4cm}

\newpage

\subsection{Heat Capacity}

The specific heat data is fitted by using equation
\begin{eqnarray} 
C_{p}(T) = mC_{vDebye}(T) + (1 - m)C_{vEinstein}(T) 
\end{eqnarray}

Where C$_{vDebye}$ is the Debye lattice heat capacity, C$_{vEinstein}$ is the Einstein lattice heat capacity. The expressions for C$_{vDebye}$ and C$_{vEinstein}$ are given by\cite{gopal2012specific}

\begin{eqnarray}
C_{vDebye}(T) =  9Nk_{B}T\bigg(\frac{T}{\theta_{D}}\bigg)^{3}\int_{0}^{\theta_{D}/T}\frac{x^{4}e^{x}}{(e^{x}-1)^{2}}dx
\end{eqnarray}

\begin{eqnarray}
	C_{vEinstein}(T) =  3NR\bigg(\frac{\theta_{E}}{T}\bigg)^{2}\frac{e^{\frac{\theta_{E}}{T}}}{(e^{\frac{\theta_{E}}{T}}-1)^{2}}dx
\end{eqnarray}

Where N is the number of atoms per formula unit, $k_{B}$ is the Boltzmann constant, $\theta_{D}$ is the Debye temperature and $\theta_{E}$ is the Einstein temperature. The magnetic part is extracted from the total heat capacity by subtracting the electronic and lattice contribution obtained from Debye and Einstein models fitting. The fitting of C$_{p}$(T) data at 5 kOe clearly shows the absence of magnetic contribution and thus results in zero magnetic entropy.

\begin{figure}[ht]
	\begin{center}
		\vspace{-5pt}
		\includegraphics[scale=0.35]{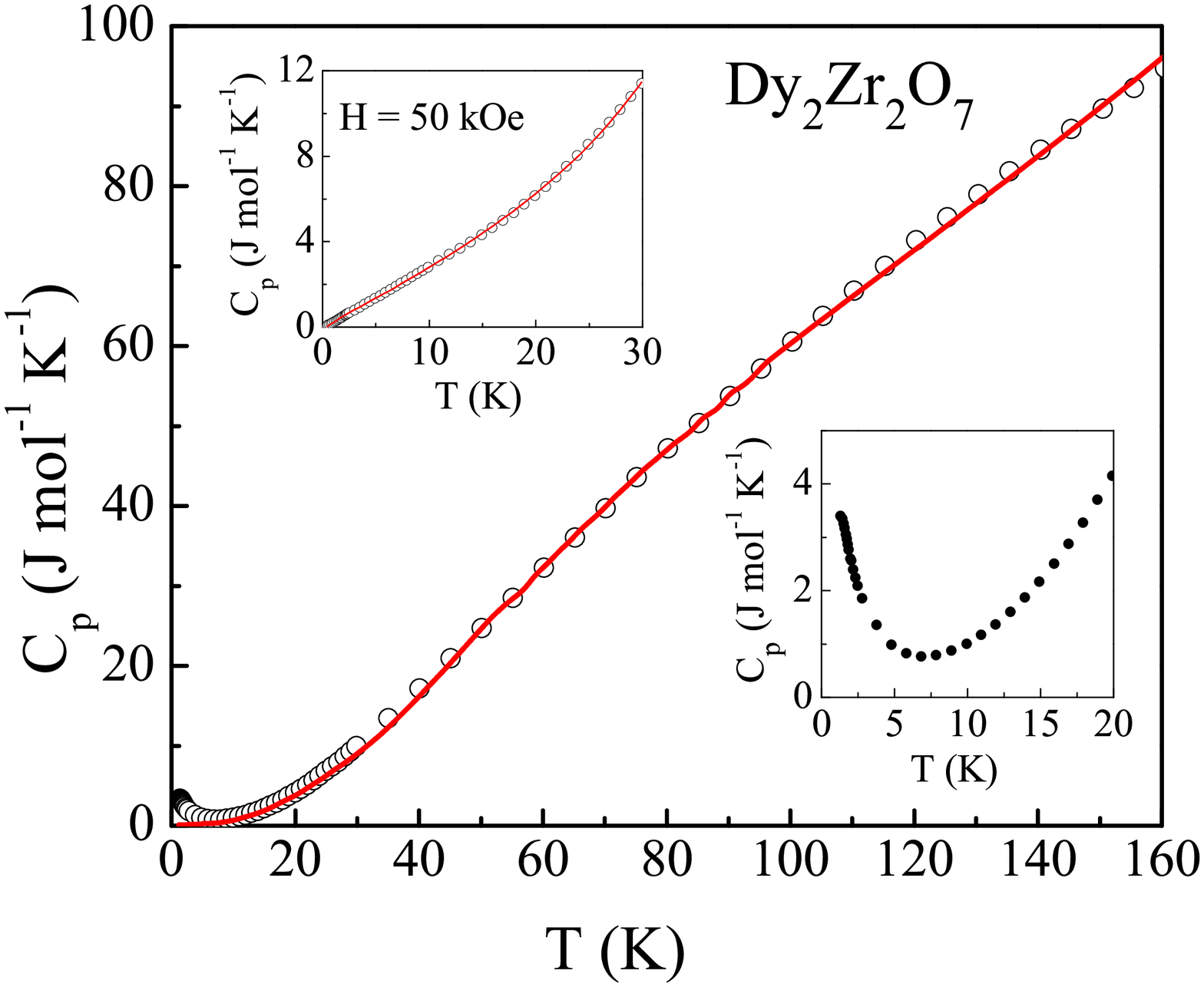}
		\vspace{-10pt}
\end{center}
\textbf{Figure S4:} (a) Temperature dependence of heat capacity C$_{p}$(T) of Dy$_{2}$Zr$_{2}$O$_{7}$ for 1.2 $\leq$ T $\leq$ 300 K measured in zero field. The red solid line is the curve fit using Debye + Einstein Models. Inset: (right) Low temperature C$_p$ for T $\leq$ 20 K. Inset: (left) Einstein and Debye models fit of heat capacity data at 50 kOe. The best fit of the data at zero field yields $\theta_{D}$ = 753(8) K, $\theta_{E}$ = 153(4) K, with the weightage of 76$\%$ for Debye term \cite{anand2015investigations}. The obtained values are in consistence with $\theta_{D}$ = 722(8) K, $\theta_{E}$ = 157(4) K for the Dy$_{2}$Ti$_{2}$O$_{7}$ \cite{anand2015investigations}. 
 \end{figure}

\end{document}